\documentclass[twoside,english]{iopart}
\usepackage[T1]{fontenc}
\usepackage[latin9]{inputenc}
\usepackage{geometry}
\usepackage{amsbsy}
\usepackage{graphicx}
\usepackage{esint}

\makeatletter

\usepackage{iopams}
\usepackage{setstack}

\usepackage[numbers, sort&compress]{natbib}
\usepackage{times}

\let\jnl@style=\rmfamily 
\def\ref@jnl#1{{\jnl@style#1}}\newcommand\physrep{\ref@jnl{Phys.~Rep.}}\newcommand\apj{\ref@jnl{ApJ}} \newcommand\mnras{\ref@jnl{MNRAS}}          \newcommand\apjl{\ref@jnl{ApJ}}          \newcommand\prd{\ref@jnl{Phys.~Rev.~D}}          
\@ifundefined{showcaptionsetup}{}{ \PassOptionsToPackage{caption=false}{subfig}}
\usepackage{subfig}
\makeatother

\usepackage{babel}
\begin{document}
\global\long\def\Mbh{M_{\bullet}}
\global\long\def\Euler{\boldsymbol{\Pi}}
\global\long\def\Tcoh{T_{\mathrm{coh}}}

\title{The statistical mechanics of relativistic orbits around a massive
black hole}

\author{{\Large{}Ben Bar-Or and Tal Alexander}}

\address{Department of Particle Physics \& Astrophysics, Faculty of Physics,
Weizmann Institute of Science, POB 26, Rehovot, Israel}

\ead{{\large{}ben.baror@weizmann.ac.il, tal.alexander@weizmann.ac.il}}
\begin{abstract}
Stars around a massive black hole (MBH) move on nearly fixed Keplerian
orbits, in a centrally-dominated potential. The random fluctuations
of the discrete stellar background cause small potential perturbations,
which accelerate the evolution of orbital angular momentum by resonant
relaxation. This drives many phenomena near MBHs, such as extreme
mass-ratio gravitational wave inspirals, the warping of accretion
disks, and the formation of exotic stellar populations. We present
here a formal statistical mechanics framework to analyze such systems,
where the background potential is described as a correlated Gaussian
noise. We derive the leading order, phase-averaged 3D stochastic Hamiltonian
equations of motion, for evolving the orbital elements of a test star,
and obtain the effective Fokker-Planck equation for a general correlated
Gaussian noise, for evolving the stellar distribution function. We
show that the evolution of angular momentum depends critically on
the temporal smoothness of the background potential fluctuations.
Smooth noise has a maximal variability frequency $\nu_{\max}$. We
show that in the presence of such noise, the evolution of the normalized
angular momentum $j=\sqrt{1-e^{2}}$ of a relativistic test star,
undergoing Schwarzschild (in-plane) General Relativistic precession
with frequency $\nu_{GR}/j^{2}$, is exponentially suppressed for
$j<j_{b}$, where $\nu_{GR}/j_{b}^{2}\sim\nu_{\max}$, due to the
adiabatic invariance of the precession against the slowly varying
random background torques. This results in an effective Schwarzschild
precession-induced barrier in angular momentum. When $j_{b}$ is large
enough, this barrier can have significant dynamical implications for
processes near the MBH.
\end{abstract}

\noindent{\it Keywords\/}: {Black holes, Galactic nuclei, Stellar dynamics and kinematics, perturbation
theory\\
}


\submitto{\CQG }

\maketitle

\makeatletter{}

\global\long\def\tw{T_{\omega}}
\global\long\def\tlin{t_{l}}
\global\long\def\Ti{T_{i}}
\global\long\def\TD{T_{D}}
\global\long\def\Tcoh{T_{\mathrm{coh}}}

\section{Introduction}

Stars around a massive black hole (MBH) move on nearly fixed Keplerian
orbits, in a centrally-dominated potential. The discrete stellar background
and its random fluctuations in time, add only small perturbations
to the MBH potential, which persist on a coherence timescale $\Tcoh$.
However, these perturbations induce a residual random torque of magnitude
$\tau_{N}\propto\sqrt{N}$, where $N$ is the number of stars on the
relevant scale, which leads to rapid evolution of the orbital angular
momentum by the process of resonant relaxation \cite{Rauch1996}.
The torques persist coherently until the background is randomized
by the slow drift away from the fixed Keplerian orbits, due for example
to the mean potential of the distributed mass (mass precession), General
Relativistic (GR) precession, or ultimately by their response to the
resonant torques themselves. Resonant relaxation is the driving force
behind many interesting physical phenomena, which include the emission
of gravitational waves from compact objects that spiral into the MBH
\cite{Amaro-Seoane2007}, the warping of circum-nuclear gaseous or
stellar disks \cite{Kocsis2011,Bregman2012}, and the orbital evolution
of tidally captured stars \cite{Perets2009,Madigan2011,Antonini2013}.
Resonant relaxation can also be a major obstacle for attempts to test
GR by observation of stars very near MBHs \cite{wil08,mer+10}. 

The original treatment of resonant relaxation \cite{Rauch1996} assumed
that a typical test orbit is similar to a typical background orbit,
and therefore there is only one relevant time scale in the system---the
coherence time $\Tcoh$ of the background. However, eccentric ($e\to1$)
relativistic orbits undergo rapid Schwarzschild (in-plane) precession
on a short timescale $(T_{GR}\propto j^{2}a^{5/2})<\Tcoh(a)$, where
$a$ is the orbital semi-major axis (sma) and $j=\sqrt{1-e^{2}}$
is the normalized orbital angular momentum. This deterministic orbital
evolution changes the torques acting on the test star over one precession
period. This motivated previous studies \cite{Rauch1996,Hopman2006}
to adopt $T_{GR}$ as the effective stochastic coherence timescale
of the random torquing process. These early studies noted that fast
torquing by resonant relaxation is expected to become inefficient
for eccentric orbits where $T_{GR}$ is much shorter than the coherence
time of the background, and where gravitational wave emission is already
dynamically significant. This would then allow inspiral by gravitational
wave emission to proceed unimpeded. However, this analysis failed
to take into account the periodic (i.e. regular rather than stochastic)
nature of the precession. In fact, depending on the statistical properties
of the background fluctuations, the effect of precession can go well
beyond merely shortening the coherence time and suppressing resonant
relaxation---it can actually lead to adiabatic invariance of the angular
momentum. Indeed, direct $N$-body simulations \cite{Merritt2011,Brem2013}
revealed that the stochastic evolution of the angular momentum is
restricted by relativistic precession at the so-called ``Schwarzschild
barrier`` \cite{Merritt2011}. However, a rigorous theoretical framework
for describing these relativistic kinematics, in terms of kinetic
and stochastic theory, is still lacking \cite{Merritt2013}.

Our goal is to study relativistic dynamics near a MBH. We focus on
the interplay between the deterministic relativistic precession and
the stochastic fluctuations of the background potential. We derive
here stochastic equations of motion (EOM), for evolving the orbital
elements of a test star, and obtain the corresponding Fokker-Planck
(FP) equation, for evolving the stellar distribution function.

In Section \ref{sec:Hamiltonian} we expand the orbit-averaged Hamiltonian
of a test star, where the effect of the background potential is represented
as a random correlated Gaussian noise. We then derive the stochastic
EOM from the leading-order stochastic Hamiltonian. The effective FP
equation for a general correlated Gaussian noise is then derived from
the EOM in Section \ref{sec:FP}. In Section \ref{sec:results} we
validate the derived FP equation by comparing its results to direct
integration of the stochastic EOM. We then show that the evolution
of the angular momentum depends not only on the values of the coherence
time $\Tcoh$ and the magnitude of the random torque $\tau_{N}$,
as expected, but also critically on the smoothness (differentiability)
of the background noise. Specifically, since a smooth background noise
has a finite maximal variability frequency $\nu_{\max}$, the evolution
of the test star's angular momentum is exponentially suppressed for
$j<j_{b}(a)$, where $T_{GR}(a,j_{b})=2\pi/\nu_{\max}(a)$. This results
in an effective barrier in angular momentum due to adiabatic invariance
induced by the relativistic precession. We argue that when $j_{b}$
is large enough to be relevant for stable relativistic orbits, this
constitutes a Schwarzschild precession-induced barrier. We discuss
the implications and summarize in section \ref{sec:Discussion}.

\makeatletter{}\global\long\def\Mbh{M_{\bullet}}
\global\long\def\Euler{\boldsymbol{\Pi}}
\global\long\def\Tcoh{T_{\mathrm{coh}}}

\section{The stochastic Hamiltonian\label{sec:Hamiltonian}}

Let $M_{\bullet}$ be the mass of the massive black hole (MBH), $M$
the mass of the test star, and $\nu_{r}=\sqrt{G\Mbh/a^{3}}$ its Keplerian
orbital frequency around the MBH, neglecting the mass of the background
stars. The star's orbital angular momentum is $J=J_{c}\sqrt{1-e^{2}}$,
where $e$ is the eccentricity and $J_{c}=\sqrt{GM_{\bullet}a}$ is
the maximal (circular) specific angular momentum for a given sma,
$a$, i.e. the star's specific binding energy in this limit is $E=G\Mbh/2a$.
The Euler angles $\Euler=\left(\psi,\theta,\phi\right)$ describing
the orbit's orientation in a fixed inertial reference frame are linearly
related to the standard orbital elements: the longitude of the ascending
node is $\Omega=\phi+\pi/2$, the inclination angle is $i=\theta$
and the argument of pericenter is $\omega=\psi-\pi/2$. Note that
neither the Euler angles nor the orbital elements are the canonical
coordinates for the Hamiltonian. 

The secular (orbit-averaged) Hamiltonian for a single star, in terms
of canonical action-angle coordinates \cite{Sridhar1999,BinneyJames2008,Touma2012}
$[(J,\psi),\,(J_{z},\phi),\,(I,w)]$, where, $J_{z}=J\cos\theta$
is the angular momentum in the $z$ direction and $I=\sqrt{G\Mbh a}$,
is 
\begin{equation}
H\left(a,J,J_{z},\phi,\psi,t\right)=H_{K}\left(a\right)+H_{GR}\left(a,J\right)+U\left(a,J,J_{z},\phi,\psi,t\right)\,,\label{eq:Htest}
\end{equation}
where due to the orbit-averaging over the mean anomaly $w$, the action
$I$, and therefore $a$, are constant. $H_{K}=-J_{c}\nu_{r}\left(a\right)/2$
is the Keplerian term, $H_{GR}=-\nu_{GR}J_{c}^{2}/J$ with $\nu_{GR}=3\nu_{r}\left(a\right)G\Mbh/\left(c^{2}a\right)$
is the relativistic 1st post-Newtonian (PN) term, and $U\left(a,J,J_{z},\phi,\psi,t\right)$
is the orbit-averaged potential due to the background stars.

\subsection{The potential of background stars}

The time-dependent, randomly fluctuating stellar background potential
can be expanded in Wigner rotation matrices $D_{mn}^{l}\left(\boldsymbol{\Pi}\right)$
(Bar-Or \& Alexander 2014, in prep.). These are used here to describe
the 3D position of an orbit, as given by its orbital elements, thereby
allowing to generalize the standard Legendre expansion of the Newtonian
potential of a collection of point masses, to that of objects that
have an additional degree of freedom. This allows the description
of both the orientation of the orbital planes, and the orientation
of the extended orbit-averaged Keplerian ellipses in them\footnote{A common use of the Wigner matrices is to describe the rotation of
point particles, where the additional degree of freedom is their intrinsic
spin.}, 
\begin{equation}
U\left(a,J,J_{z},\phi,\psi,t\right)=\sum_{l=0}^{\infty}H_{l}\left(a,J,J_{z},\phi,\psi,t\right),\label{eq:Potential}
\end{equation}
where the $l$'th degree spherical harmonic is further decomposed
to $(2l+1)^{2}$ multipoles  
\begin{eqnarray}
H_{l}\left(a,J,J_{z},\phi,\psi,t\right) & = & \sum_{m=-l}^{l}\sum_{n=-l}^{l}D_{nm}^{l}\left(\boldsymbol{\Euler}\right)h_{nm}^{l}\left(J,a,t\right)\,,\label{eq:Hl}
\end{eqnarray}
and where the terms 
\begin{eqnarray}
h_{nm}^{l}\left(a,J,t\right) & = & -\tau_{0}\sum_{k=1}^{N}\frac{4\pi}{2l+1}\sum_{n_{k}=-l}^{l}D_{n_{k}m}^{l}\left(\boldsymbol{\Euler}_{k}(t)\right)^{*}\nonumber \\
 &  & \times Y_{l}^{n{}_{k}}\left(\frac{\pi}{2},0\right)Y_{l}^{n}\left(\frac{\pi}{2},0\right)^{*}\left\langle \frac{a}{r}K_{l}\left(r_{k}\left[a_{k}(t),e_{k}(t),f_{k}\right]/r\right)e^{i\left(n_{k}f_{k}-nf\right)}\right\rangle _{\circlearrowright}\,.\label{eq:etalnm}
\end{eqnarray}
depend on the distribution of the stellar background, which is a function
of time through the evolution of the orbital elements of the $N$
stars, and is represented below by a noise model. Note that for each
$(l,n,m)$ term, the time-dependence of the stellar background enters
only through $h_{nm}^{l}$. The scale of the torque $\tau_{0}\left(a\right)=J_{c}\nu_{r}Q^{-1}$
due to a single background star is inversely proportional to the MBH-star
mass ratio $Q=\Mbh/M$ (a single-mass population assumed). The usual
$\left(r_{<}/r_{>}\right)^{l}/r_{>}$ min-max terms that appear in
the expansion of the potential to Legendre polynomials are rewritten
in terms of the dimensionless functions $K_{l}\left(x\right)\equiv\min\left(x,1\right)^{2l+1}/x^{l+1}.$
The average $\left\langle \dots\right\rangle _{\circlearrowright}$
denotes orbit-averaging over the mean anomalies of both the test star
and the $k$'th background star.

\subsection{The background potential in the stochastic limit}

A key assumption is that correlations between any two stars in the
system are short-lived, and have a negligible affect on the long-term
evolution of the system.  Since each $h_{nm}^{l}\left(a,J,t\right)$
term results from the superposed gravitational forces by $N\gg1$
background stars, we invoke the central limit theorem and assume that
it can be described by time-dependent Gaussian random variables $\eta_{nm}^{l}\equiv h_{nm}^{l}-\left\langle h_{nm}^{l}\right\rangle $,
which have zero mean and are therefore completely described by their
2-point correlation functions (Eq. \ref{eq:Chl}).  Therefore, although
the system is completely deterministic, the time-dependent terms in
the Hamiltonian can be considered as a time-correlated background
``noise''. The derived EOM can then be interpreted as a set of non-linear
Langevin equations (e.g. \cite[Ch. 3]{Risken1989}).

We assume that the stellar background is on average isotropic and
stationary, so that only the spherical monopole component ($l=0$)
has a non-zero mean, 

\begin{equation}
\left\langle h_{00}^{0}\left(a,J,t\right)\right\rangle =4\pi N\tau_{0}\int n\left(a_{k},e_{k}\right)da_{k}de_{k}\left\langle \frac{a}{r}K_{0}\left(r_{k}/r\right)\right\rangle _{\circlearrowright}\,,\label{eq:meanh0}
\end{equation}
\begin{equation}
\left\langle h_{nm}^{l}\left(a,J,t\right)\right\rangle =0,\; l>0\,,\label{eq:meanhl}
\end{equation}
where $\left\langle \ldots\right\rangle $ denotes the mean over all
initial conditions and realizations of the background, and $n(a,e)$
is the number density of stars in $(a,e)$ phase space. The 2-point
correlation functions $C_{n,n^{\prime}}^{l}$ are then simplified
by the orthonormality of the Wigner $D$-matrices (Eq. \ref{eq:etalnm})

\begin{eqnarray}
\left\langle \eta_{nm}^{l}\left(a,J,t\right)\eta_{n^{\prime}m^{\prime}}^{l^{\prime}*}\left(a^{\prime},J^{\prime},t^{\prime}\right)\right\rangle  & = & \delta_{ll^{\prime}}\delta_{mm^{\prime}}C{}_{nn^{\prime}}^{l}\left(a,a^{\prime},J,J^{\prime},t-t^{\prime}\right)\,.\label{eq:Chl}
\end{eqnarray}
This expansion of the Hamiltonian into $(l,n,m)$ terms has the useful
property of decoupling the angular dependence of the test orbit from
that of the background for each term separately (Eq. \ref{eq:Hl}).
Note that since the auto-correlations $\left\langle \eta_{nm}^{l}\left(a,J,t\right)\eta_{nm}^{l^{\prime}*}\left(a,J,t\right)\right\rangle $
are independent of $m$, all orders of $m$ for a given $l$ and $n$
contribute equally to the Hamiltonian and should be taken into account.

\subsection{First order ($l=1$) Hamiltonian}

The $l\le1$ terms in the Hamiltonian are 
\begin{eqnarray}
H\left(a,J,J_{z},\phi,t\right) & = & -J_{c}\nu_{r}\left(a\right)/2-\nu_{GR}J_{c}^{2}/J+h_{00}^{0}\left(J,a,t\right)\nonumber \\
 &  & +\sum_{m=-1}^{1}\sum_{n=-1}^{1}D_{nm}^{1}\left(\boldsymbol{\Pi}\right)h_{nm}^{1}\left(J,a,t\right)\,,\label{eq:1st-H}
\end{eqnarray}
In general the $l=1$ term of the Hamiltonian has $\left(2l+1\right)^{2}=9$
real-valued $h_{nm}^{1}$ components, which correspond to 9 real-valued
noise components\footnote{The $h_{00}^{0}$ term, which contributes only to the evolution of
$\psi$ (precession due to the distributed stellar mass), is the only
one with a non-zero mean. The mean $\left\langle h_{00}^{0}\right\rangle $
leads to a constant drift in $\psi$, while the noise associated with
this term, $\eta_{00}^{0}=h_{00}^{0}-\left\langle h_{00}^{0}\right\rangle $
(i.e. fluctuations in the total stellar mass within a fixed radius),
is a higher-order stochastic perturbation.}. However, the symmetries of $h_{nm}^{1}$ imply that $\eta_{0m}^{1}=0$
and $\eta_{-1m}^{1}=\eta_{1m}^{1}$, leaving only three non-zero noise
terms. We can therefore decompose the noise into three independent
real Gaussian components $\boldsymbol{\tilde{\eta}}\left(a,J,t\right)=\left(\eta_{1}\left(a,J,t\right),\eta_{2}\left(a,J,t\right),\eta_{3}\left(a,J,t\right)\right)^{T}$
\begin{eqnarray}
\eta_{\pm11}^{1}\left(a,J,t\right) & = & \frac{1}{\sqrt{2}}\left(\tilde{\eta}_{1}\left(a,J,t\right)+i\tilde{\eta}_{2}\left(a,J,t\right)\right)\,,\\
\eta_{\pm1-1}^{1}\left(a,J,t\right) & = & -\frac{1}{\sqrt{2}}\left(\tilde{\eta}_{1}\left(a,J,t\right)-i\tilde{\eta}_{2}\left(a,J,t\right)\right)\,,\\
\eta_{\pm10}^{1}\left(a,J,t\right) & = & \tilde{\eta}_{3}\left(a,J,t\right)\,.
\end{eqnarray}

Note that the noise components $\tilde{\eta}_{i}$ depend on the position
of the test orbit in $(a,J)$ phase-space, and likewise, so do the
$2$-point correlation functions. Here we simplify this general description
by assuming that on the relevant timescales, which are shorter than
the 2-body energy relaxation timescale, $a$ remains nearly constant,
so the coherence scales can be evaluated at the fixed, initial $a$
value.

We further assume that at any given short time interval $(t,t+\mathrm{d}t)$
over which the test orbit's angular momentum evolves over the interval
$(J,J+\mathrm{d}J)$, the temporal correlations decay much faster
than the angular momentum correlations, $\left\langle \eta\left(J,t\right)\eta\left(J,t+dt\right)\right\rangle \ll\left\langle \eta\left(J,t\right)\eta\left(J+dJ,t\right)\right\rangle $,
where $dJ\sim\tau_{N}dt$. This assumption is motivated by the fact
that $\Tcoh<J_{c}/\tau_{N}$, whether the coherence time is determined
by mass precession (e.g. \cite{Eilon2009}), or whether it is determined
by the resonant torques themselves (as indicated by fully self-consistent
numerical evolution simulations of circular orbits, Bar-Or \& Alexander
2014, in prep.). This allows the separation $\boldsymbol{\tilde{\eta}}\left(a,J,t\right)\approx\sqrt{C^{1}\left(a,J\right)}\boldsymbol{\eta}\left(t\right)$,
where the amplitude $C^{1}\left(a,J\right)=C_{11}^{1}\left(a,a,J,J,0\right)$
is normalized to reproduce the covariance matrix, and where $\boldsymbol{\eta}$
is a vector with three independent Gaussian components, with zero
mean and an auto-correlation function (ACF) 
\begin{equation}
\left\langle \eta_{i}\left(t\right)\eta_{j}\left(t^{\prime}\right)\right\rangle =\delta_{ij}C\left(t-t^{\prime}\right),\label{eq:Chh}
\end{equation}
where $C(0)=1$. The ACF is characterized by its shape and magnitude.
We define here the coherence time as 
\begin{equation}
\Tcoh=\int_{0}^{\infty}C\left(t\right)dt\,.\label{eq:Tcoh}
\end{equation}
Note that as defined, $\Tcoh$ is half the total power of the noise. 

The assumption of separability leads to a great simplification of
the stochastic EOM, since the noise term $\boldsymbol{\eta}$ then
enters as function of time only, without derivatives of the stochastic
noise with respect to $J$. The $l\le1$ stochastic Hamiltonian then
reads 
\begin{equation}
H_{\eta}=-\nu_{GR}J_{c}^{2}/J+\left\langle h_{00}^{0}\left(J,a,t\right)\right\rangle +\eta_{00}^{0}\left(J,a,t\right)+\sqrt{2C^{1}\left(a,J\right)}\hat{e}_{\psi}(\boldsymbol{\Pi})\cdot\boldsymbol{\eta}\left(t\right),\label{eq:Heta3}
\end{equation}
where 
\begin{equation}
\hat{e}_{\psi}(\boldsymbol{\Pi})\equiv\sin\psi\hat{e}_{\phi}(\phi)-\cos\psi\hat{e}_{u}(\phi,u)\,.
\end{equation}
and we introduce an orthonormal spherical coordinate system $\left(J,\phi,u\right)$,
where $u\equiv\cos\theta$, with the associated unit vectors $\hat{e}_{i}=\left(\partial\mathbf{J}/\partial i\right)/\left|\partial\mathbf{J}/\partial i\right|$
for $i\in\left\{ J,\phi,u\right\} $
\begin{equation}
\hat{e}_{J}=\left(\begin{array}{c}
\sqrt{1-u^{2}}\cos\phi\\
\sqrt{1-u^{2}}\sin\phi\\
u
\end{array}\right);\;\hat{e}_{\phi}=\left(\begin{array}{c}
-\sin\phi\\
\cos\phi\\
0
\end{array}\right);\;\hat{e}_{u}=\left(\begin{array}{c}
-u\cos\phi\\
-u\sin\phi\\
\sqrt{1-u^{2}}
\end{array}\right),
\end{equation}
 The noise term $\boldsymbol{\eta}$ can be viewed as a 3D vector
in angular momentum space. In this sense, $\boldsymbol{\eta}$ plays
a role similar to the ``dipole'' term in the Hamiltonian model of
Merritt et al. \cite{Merritt2011} (see Eq. A6 there), which used
the \emph{ansatz} of an effective potential to represent the residual
torque by the background stars. Here, in contrast, we derive the Hamiltonian
from first principles without invoking an effective potential, and
include the stochastic effect of the background stars as a formally
defined noise term.

\subsection{Integrating the stochastic equations of motion}

\label{sub:intEOM}

We are interested here only in the coupling between the GR precession
and the stochastic Newtonian perturbations of the background stars
on the test orbit, we therefore omit the Newtonian spherical terms
(mass precession terms), and $H_{\eta}$ reduces to 

\begin{equation}
H_{\eta}^{GR}=-\nu_{GR}J_{c}/j+\tau_{N}^{1}(j)\hat{e}_{\psi}(\boldsymbol{\Pi})\cdot\boldsymbol{\eta}\left(t\right),\label{eq:HetaGR}
\end{equation}
where $\tau_{N}^{1}=\sqrt{2C^{1}\left(a,j\right)}\sim\sqrt{N(a)}\tau_{0}$
is the scale of the residual torque, which reflects the random contribution
of the $N\left(a\right)$ stars on the relevant scale. Using static
wires simulation similar to \cite{Gurkan2007} but with a faster method
\cite{Touma2009}, we calculated the total residual torque $\tau_{N}$
on a test wire in the $ $$\hat{e}_{J}$ direction due to a cluster
of $N$-wires. The total torque can be approximated by $\tau_{N}\left(a,j\right)=\sqrt{\left\langle \tau_{J}^{2}\right\rangle }\approx0.28\sqrt{1-j}\sqrt{N\left(2a\right)}GM/a$.
This is larger than the partial torque $\tau_{N}^{1}$ corresponding
to the $l=1$ term, which rises asymptotically to $\approx0.6\tau_{N}\left(a,j=0\right)$
as $j\to0$ (by numeric integration of $C_{1,1}^{1}$, using Eq. \ref{eq:etalnm}).
Nevertheless, we replace $\tau_{N}^{1}$ with $\tau_{N}$. This has
the advantage of simplicity, and of reproducing the total power of
the residual torques with the $l=1$ approximation, which may be a
better model of the real physical system.

The EOM derived from $H_{\eta}^{GR}$ can be reduced to the compact
form 

\begin{eqnarray}
\dot{\boldsymbol{J}} & = & -\tau_{N}\left(j\right)\hat{e}_{\psi}\left(\boldsymbol{\Pi}\right)\times\boldsymbol{\eta}\left(t\right),\label{eq:EOMJ}\\
\dot{\psi} & = & \nu_{GR}j^{-2}+\frac{\tau_{N}\left(j\right)}{J}\left[\frac{\partial\log\tau_{N}\left(j\right)}{\partial\log J}\hat{e}_{\psi}\left(\boldsymbol{\Pi}\right)-\frac{J_{z}\cos\psi}{\sqrt{J^{2}-J_{z}^{2}}}\hat{e}_{J}\left(\boldsymbol{\Pi}\right)\right]\cdot\boldsymbol{\eta}\left(t\right)\label{eq:EOMpsi}
\end{eqnarray}
 In this form the EOM are non-linear Langevin equations, which require
a specified \emph{noise model} $\boldsymbol{\eta}(t)$. The system
can be integrated in time from specified initial conditions, for any
given realization of the noise. The statistical properties of the
system can then be obtained by repeating this for many noise realizations,
in a Monte Carlo fashion. Alternatively, the evolution of the probability
density function (PDF) of the test star in phase-space can be derived
from the corresponding FP equation. This is a well-defined procedure
in the Markovian limit, where $\boldsymbol{\eta}$ can be treated
as an uncorrelated white noise on all physically relevant timescales.
In the general non-Markovian case, it is not always possible to describe
the dynamics by an effective FP equation. However, the system of interest
here can be approximately described in this form (Section \ref{sec:FP}),
as is validated by comparison to direct integration of the stochastic
EOM (Section \ref{sec:results}).

\makeatletter{}\global\long\def\Mbh{M_{\bullet}}
\global\long\def\Euler{\boldsymbol{\Pi}}
\global\long\def\Tcoh{T_{\mathrm{coh}}}

\section{Effective diffusion with correlated noise}

\label{sec:FP}

In the $N\gg1$ limit, the fluctuating potential of the background
stars can be treated as a stochastic noise. The EOM for $\phi$, $u=\cos\theta$,
$\psi$ and total angular momentum $j=J/J_{c}$ are then derived from
the stochastic Hamiltonian $H_{\eta}$ (Eq. \ref{eq:Heta3}), 
\begin{eqnarray}
\dot{x} & = & \boldsymbol{\nu}_{x}\left(j,\boldsymbol{\Pi}\right)\cdot\boldsymbol{\eta}\left(t\right);\;\; x\in\left\{ j,\phi,u\right\} ,\label{eq:xdot}\\
\dot{\psi} & = & \nu_{p}\left(j\right)+\boldsymbol{\nu}_{\psi}\left(j,\boldsymbol{\Pi}\right)\cdot\boldsymbol{\eta}\left(t\right).\label{eq:psidot}
\end{eqnarray}
where the vectors $\boldsymbol{\nu}_{x}\left(j,\boldsymbol{\Pi}\right)$
express the torques in the $(j,\phi,u,\omega)$ coordinates, and transform
the noise $\boldsymbol{\eta}$ to them from the $\mathbf{J}$ phase-space
in which it is defined. 
\begin{eqnarray}
\boldsymbol{\nu}_{j}\left(j,\boldsymbol{\Pi}\right) & = & -\nu_{j}\left(j\right)\hat{e}_{\psi\psi}\left(\boldsymbol{\Pi}\right),\label{eq:nuj}\\
\boldsymbol{\nu}_{\phi}\left(j,\boldsymbol{\Pi}\right) & = & \nu_{j}\left(j\right)j^{-1}\frac{\cos\psi}{\sqrt{1-u^{2}}}\hat{e}_{J}\left(\boldsymbol{\Pi}\right),\label{eq:nuphi}\\
\boldsymbol{\nu}_{u}\left(j,\boldsymbol{\Pi}\right) & = & \nu_{j}\left(j\right)j^{-1}\sin\psi\sqrt{1-u^{2}}\hat{e}_{J}\left(\boldsymbol{\Pi}\right),\label{eq:nuu}\\
\boldsymbol{\nu}_{\psi}\left(j,\boldsymbol{\Pi}\right) & = & \frac{\partial\nu_{j}\left(j\right)}{\partial j}\hat{e}_{\psi}\left(\boldsymbol{\Pi}\right)-\frac{\nu_{j}\left(j\right)}{j}\frac{u\cos\psi}{\sqrt{1-u^{2}}}\hat{e}_{J}\left(\boldsymbol{\Pi}\right),\label{eq:nupsi}
\end{eqnarray}
with $\nu_{j}\left(j\right)=\left|\boldsymbol{\mathbf{\nu}}_{j}\left(j\right)\right|=\tau_{N}\left(J\right)/J_{c}$
and $\hat{e}_{\psi\psi}=\partial\hat{e}_{\psi}/\partial\psi$.

These EOM are a set of non-linear Langevin type equations. The corresponding
FP equations, which describe the evolution of the PDF of the state
variables, $P\left(j,\phi,u,\psi,t\right)$, can be derived in the
Markovian limit (e.g. \cite{Risken1989}). For simplicity, we focus
here only on the PDF of the normalized angular momentum, $j$, marginalizing
over all the other state variables. 

For a \emph{given} realization of the noise, the trajectory of a test
star in phase-space, $(j(t),\boldsymbol{\Pi}(t)),$ is fully specified
by Eqs. (\ref{eq:xdot}, \ref{eq:psidot}) and the initial conditions.
Therefore, its $j$-trajectory can be formally described by $\varphi\left(j,t\right)=\delta\left(j-j\left(t\right)\right)$,
which obeys the continuity equation

\begin{eqnarray}
\frac{\partial}{\partial t}\varphi\left(j,t\right) & =-\frac{\partial}{\partial j}\left[\frac{dj}{dt}\left(j,t\right)\varphi\left(j,t\right)\right]= & -\frac{\partial}{\partial j}\left[\boldsymbol{\nu}_{j}\left(j,\boldsymbol{\Pi}\left(t\right)\right)\cdot\boldsymbol{\eta}\left(t\right)\varphi\left(j,t\right)\right].
\end{eqnarray}
The PDF is then given by $P\left(j,t\right)=\left\langle \varphi\left(j,t\right)\right\rangle $,
where $\left\langle \dots\right\rangle $ denotes the average over
all realizations and all initial conditions of the noise. The evolution
of $P\left(j,t\right)$ is determined by
\begin{equation}
\frac{\partial}{\partial t}P\left(j,t\right)=-\frac{\partial}{\partial j}\left\langle \boldsymbol{\nu}_{j}\left(j,\boldsymbol{\Pi}\left(t\right)\right)\cdot\boldsymbol{\eta}\left(t\right)\varphi\left(j,t\right)\right\rangle .\label{eq:Pdot_cont}
\end{equation}

Since the noise term $\boldsymbol{\eta}\left(t\right)$ is a Gaussian
with zero mean, Novikov's theorem \cite{Novikov1965}, which expresses
the correlation of the noise with any functional $R\left[\boldsymbol{\eta}\right]$
of the noise, can be applied, 
\begin{equation}
\left\langle \eta_{i}\left(t\right)R\left[\boldsymbol{\eta}(t^{\prime})\right]\right\rangle =\int ds\left\langle \eta_{i}\left(t\right)\eta_{j}\left(s\right)\right\rangle \left\langle \frac{\delta R\left[\boldsymbol{\eta}(t^{\prime})\right]}{\delta\eta_{j}\left(s\right)}\right\rangle \,.
\end{equation}
Therefore, Eqs. (\ref{eq:Chh}) and (\ref{eq:Pdot_cont}) imply

\begin{eqnarray}
\frac{\partial}{\partial t}P\left(j,t\right) & = & -\frac{\partial}{\partial j}\left\{ \int_{0}^{t}dt^{\prime}C\left(t-t^{\prime}\right)\left\langle \nabla_{\eta}\left(t^{\prime}\right)\cdot\left[\boldsymbol{\nu}_{j}\left(j,\boldsymbol{\Pi}\left(t\right)\right)\varphi\left(j,t\right)\right]\right\rangle \right\} \,,\label{eq:P_dot_Nov}
\end{eqnarray}
where we define for brevity the functional gradient operator 
\begin{equation}
\nabla_{\eta}\left(t\right)\equiv\left(\frac{\delta}{\delta\eta_{1}\left(t\right)},\frac{\delta}{\delta\eta_{2}\left(t\right)},\frac{\delta}{\delta\eta_{3}\left(t\right)}\right)\,.
\end{equation}
By multiple application of the chain rule, Eq. (\ref{eq:P_dot_Nov})
reads
\begin{eqnarray}
\frac{\partial}{\partial t}P\left(j,t\right) & = & \frac{\partial^{2}}{\partial j^{2}}\left\{ \int_{0}^{t}dt^{\prime}C\left(t-t^{\prime}\right)\left\langle \varphi\left(j,t\right)\boldsymbol{\nu}_{j}\left(j,\boldsymbol{\Pi}\left(t\right)\right)\cdot\nabla_{\eta}\left(t^{\prime}\right)j\left(t\right)\right\rangle \right\} \nonumber \\
 &  & -\frac{\partial}{\partial j}\left\{ \int_{0}^{t}dt^{\prime}C\left(t-t^{\prime}\right)\sum_{x=\left\{ j,\phi,u,\psi\right\} }\left\langle \varphi\left(j,t\right)\frac{\partial\boldsymbol{\nu}_{j}\left(j,\boldsymbol{\Pi}\left(t\right)\right)}{\partial x}\cdot\nabla_{\eta}\left(t^{\prime}\right)x\left(t\right)\right\rangle \right\} \nonumber \\
\label{eq:P_dot_Nov2}
\end{eqnarray}
We integrate the EOM (Eqs. \ref{eq:xdot}, \ref{eq:psidot}) to obtain
\begin{eqnarray}
x\left(t\right) & = & x\left(0\right)+\int_{0}^{t}ds\boldsymbol{\nu}_{x}\left(j\left(s\right),\boldsymbol{\Pi}\left(s\right)\right)\cdot\boldsymbol{\eta}\left(s\right);\;\; x\in\left\{ j,\phi,u\right\} ,\\
\psi\left(t\right) & = & \psi\left(0\right)+\int_{0}^{t}\nu_{p}\left(j\left(s\right)\right)ds+\int_{0}^{t}\boldsymbol{\nu}_{\psi}\left(j\left(s\right),\boldsymbol{\Pi}\left(s\right)\right)\cdot\boldsymbol{\eta}\left(s\right).
\end{eqnarray}
Thus, for $t^{\prime}\le t$, the response functions for $x\in\left\{ j,\phi,u,\psi\right\} $
are
\begin{eqnarray}
\nabla_{\eta}\left(t^{\prime}\right)x\left(t\right) & = & \boldsymbol{\nu}_{x}\left(j\left(t^{\prime}\right),\boldsymbol{\Pi}\left(t^{\prime}\right)\right)\nonumber \\
 &  & +\sum_{y\in\left\{ j,\phi,u,\psi\right\} }\int_{t^{\prime}}^{t}\frac{\partial\boldsymbol{\nu}_{x}}{\partial y}\left(j\left(s\right),\boldsymbol{\Pi}\left(s\right)\right)\cdot\boldsymbol{\eta}\left(s\right)\nabla_{\eta}\left(t^{\prime}\right)y\left(s\right)ds\,.\label{eq:Delhx}
\end{eqnarray}
These functions describe the manner by which a small perturbation
of the noise at an initial time $t^{\prime}$ is magnified at a later
time $t$ by the accumulated differences in torques experienced along
the perturbed and unperturbed trajectories, due to the fact that the
noise amplitude ($\boldsymbol{\nu}_{x}$) depends on the state variables
along the trajectory. 

In general, Eq. (\ref{eq:P_dot_Nov2}) can be written in closed form
(i.e. solely in terms of $P$) only when the response of the state
variables to the noise (given by the response functions $\nabla_{\eta}\left(t^{\prime}\right)x\left(t\right)$)
can be reduced to a form that involves only deterministic functions
(i.e. a known function of $j(t),$ $t$ and $t^{\prime}$). Otherwise,
the Novikov theorem has to be applied recursively, but closure is
not guaranteed \cite{Sancho1989,Hanggi1995}. 

To address this difficulty, we first consider the Markovian limit,
where this problem is avoided. In this limit, the process depends
only on the total power of the noise, and therefore can be presented
by an appropriately normalized uncorrelated noise. Eq. (\ref{eq:P_dot_Nov2})
then reduces to the FP equation (section \ref{sub:Markov}). We further
show below, that in the non-Markovian case, which is relevant in the
relativistic limit where the precession time is short, the response
functions can be approximated by deterministic functions, which we
then employ to convert Eq. (\ref{eq:P_dot_Nov2}) to an effective
FP equation (section \ref{sub:nonMarkov}).

\subsection{The Markovian Limit}

\label{sub:Markov}

In the limit where the correlation time is shorter than any other
relevant timescale in the system, the noise can be regarded as an
uncorrelated (white) noise with a correlation function $C\left(|t-t^{\prime}|\right)=2\Tcoh\delta\left(t-t^{\prime}\right)$,
and the process is Markovian\footnote{Note that in this singular case, $C(0)\to\infty$ is no longer normalized
to $C(0)=1$ as previously assumed. This inconsistency is not a problem,
because in this limit the dynamics depend only on the total power
of the noise, $2\Tcoh$ (Eq. \ref{eq:Tcoh}), through the diffusion
coefficient $D_{2}=2\tau_{N}^{2}\Tcoh$.}. In this limit, the response functions contribute to the integrals
in Eq. (\ref{eq:P_dot_Nov2}) only at the limit $t^{\prime}=t$ and
therefore 
\begin{eqnarray}
\frac{\partial}{\partial t}P\left(j,t\right) & = & \Tcoh\frac{\partial^{2}}{\partial j^{2}}\left\{ \nu_{j}^{2}\left(j\right)P\left(j,t\right)\right\} \nonumber \\
 &  & -\Tcoh\sum_{x=\left\{ j,\phi,u,\psi\right\} }\frac{\partial}{\partial j}\left\{ \left\langle \frac{\partial\boldsymbol{\nu}_{j}\left(j,\boldsymbol{\Pi}\left(t\right)\right)}{\partial x}\cdot\boldsymbol{\nu}_{x}\left(j,\boldsymbol{\Pi}\left(t\right)\right)\varphi\left(j,t\right)\right\rangle \right\} \,,
\end{eqnarray}
where we used 
\begin{eqnarray}
\left.\nabla_{\eta}\left(t^{\prime}\right)x\left(t\right)\right|_{t\to t^{\prime}} & = & \boldsymbol{\nu}_{x}\left(j\left(t\right),\boldsymbol{\Pi}\left(t\right)\right);\;\; x\in\left\{ j,\phi,u,\psi\right\} .
\end{eqnarray}
It then follows from
\begin{eqnarray}
\sum_{x=\left\{ j,\phi,u,\psi\right\} }\frac{\partial\boldsymbol{\nu}_{j}\left(j,\boldsymbol{\Pi}\left(t\right)\right)}{\partial x}\cdot\boldsymbol{\nu}_{x}\left(j\left(t\right),\boldsymbol{\Pi}\left(t\right)\right) & = & \frac{1}{j}\frac{\partial}{\partial j}j\nu_{j}^{2}\left(j\right),
\end{eqnarray}
that the FP equation is

\begin{eqnarray}
\frac{\partial}{\partial t}P\left(j,t\right) & = & \frac{1}{2}\frac{\partial}{\partial j}\left\{ jD_{2}(j)\frac{\partial}{\partial j}\left[\frac{1}{j}P\left(j,t\right)\right]\right\} ,\label{eq:FP}
\end{eqnarray}
where $D_{2}(j)=D(\Delta j^{2})=2\Tcoh\nu_{j}^{2}\left(j\right)$
is the usual 2nd order diffusion coefficient (DC). It is easy to verify
that the zero-flux steady state satisfies the maximal entropy solution,
$P(j)=2j$, as it should \cite{Rauch1996}, for any positive $D_{2}$.
This FP equation can be written in standard form using the 1st order
DC $D_{1}(j)=D(\Delta j)$, as $\partial P/\partial t=(1/2)\partial^{2}D_{2}P/2\partial j^{2}-\partial D_{1}P/\partial j$,
where the two DCs are related by $2jD_{1}\left(j\right)=\partial\left(jD_{2}\left(j\right)\right)/\partial j$,
as can also be derived by substituting the maximal entropy solution.

Note that the EOM do not contain a true drift term in $j$. However,
a ``parametric drift'' \cite{Risken1989} arises because of the
gradients in the noise amplitude terms $\boldsymbol{\nu}_{x}(j,\boldsymbol{\Pi})$
(Eqs. \ref{eq:xdot}, \ref{eq:psidot}), which imply that consecutive
random steps in the direction of increasing amplitude will tend to
be larger than in the opposite direction, even though each individual
step is intrinsically symmetric.

\subsection{The non-Markovian case}

\label{sub:nonMarkov}

In practice, the Markovian limit is not applicable for the dynamics
considered here, since the ACF has a finite correlation time $\Tcoh$,
which can be much longer than the precession period $T_{p}=2\pi/\nu_{p}\left(j\right)$.

The analysis can be extended using several assumptions, which are
verified numerically below (Section \ref{sec:results}). Since the
decay timescale of $C\left(t-t^{\prime}\right)$ is $\Tcoh$, the
response functions mostly contribute to the integral in Eq. (\ref{eq:P_dot_Nov2})
at $|t-t^{\prime}|\ll\Tcoh<\nu_{j}^{-1}$. Thus, the second term in
Eq. (\ref{eq:Delhx}) is of the order of $\nu_{j}^{2}|t-t^{\prime}|\ll\Tcoh\nu_{j}^{2}<\nu_{j}$,
which is much smaller than the first term ( $\sim\nu_{j}$), and can
therefore be neglected. Moreover, on timescales $|t-t^{\prime}|\ll\Tcoh<\nu_{j}^{-1}$,
the state variables $j$, $\phi$ and $u$ (which evolve stochastically
on the timescale $\nu_{j}^{-1}$) are almost constant, while $\psi$
evolves deterministically (linearly) in time as $\nu_{p}\left(j\right)(t-t^{\prime})$.
Therefore, the response functions can be approximated as

\begin{eqnarray}
\nabla_{\eta}\left(t^{\prime}\right)x\left(t\right) & \approx & \boldsymbol{\nu}_{x}\left[j\left(t\right),\phi\left(t\right),u\left(t\right),\psi\left(t\right)+\nu_{p}\left(j\left(t\right)\right)(t-t^{\prime})\right];\;\; x\in\left\{ j,\phi,u,\psi\right\} .
\end{eqnarray}
This gives rise to an effective FP equation 
\begin{eqnarray}
\frac{\partial}{\partial t}P\left(j,t\right) & = & \frac{1}{2}\frac{\partial}{\partial j}\left\{ jD_{2}\left(j\right)\frac{\partial}{\partial j}\left[\frac{1}{j}P\left(j,t\right)\right]\right\} \,,\label{eq:EffFP}
\end{eqnarray}
with an effective DC  
\begin{eqnarray}
D_{2}\left(j\right) & = & 2\nu_{j}^{2}\left(j\right)\int_{0}^{\infty}\mathrm{d}tC\left(t\right)\cos\left(\nu_{p}\left(j\right)t\right)=\nu_{j}^{2}\left(j\right)S_{\eta}\left(\nu_{p}\left(j\right)\right),\label{eq:D_a}
\end{eqnarray}
where $S_{\eta}\left(\omega\right)=\mathcal{F}\left[C\left(t\right)\right](\omega)$
is the spectral power density of the noise and $\mathcal{F}$ is the
Fourier transform. Since the treatment is only applicable for $t\gg\Tcoh$,
we took the integration limit in Eq. (\ref{eq:P_dot_Nov2}) to infinity.

Note that in the limit where there is no precession i.e. $\nu_{p}\to0$,
the diffusion coefficient approaches the Markovian limit $D_{2}\left(j\right)=2\nu_{j}^{2}\left(j\right)\Tcoh$.
Previous studies (e.g. \cite{Hopman2006}) identified the diffusion
time associated with resonant relaxation as $T_{RR}\sim J_{c}^{2}/\left(\tau_{N}^{2}\Tcoh\right)\sim1/D_{2}$,
which is equivalent to assuming the Markovian limit. This is a valid
approximation in the Newtonian case when $\nu_{GR}/j^{2}<2\pi/\Tcoh$
\footnote{It is assumed here that mass precession does not induce adiabatic
invariance. This  remains to be studied in detail.}.

\begin{figure}
\subfloat[]{\includegraphics[width=0.5\textwidth]{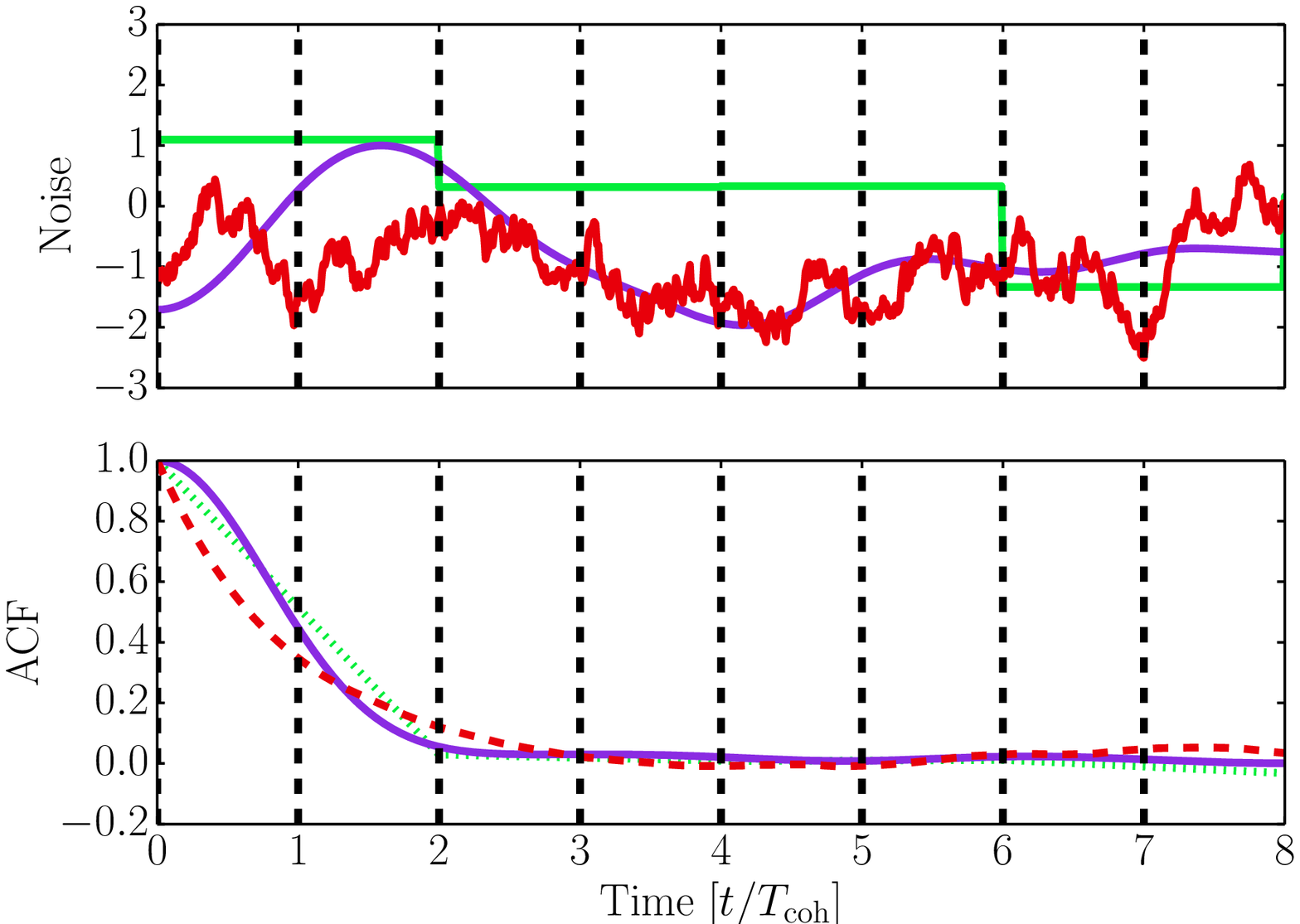}}\subfloat[]{\includegraphics[width=0.44\textwidth]{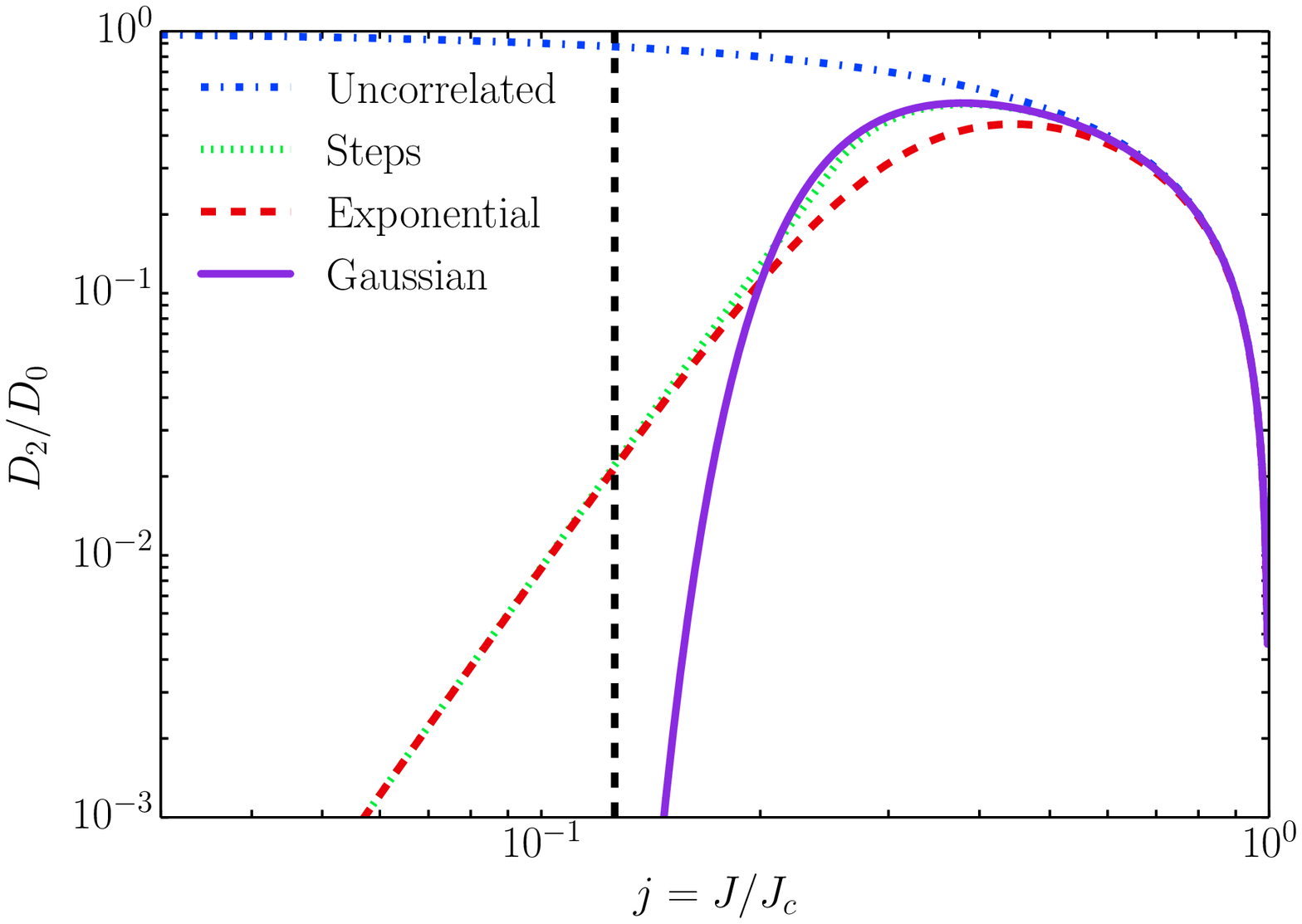}}

\protect\caption{\label{fig:DCs}a) Different correlated noise models (upper panel),
and their auto-correlation functions (lower panel). b) The normalized
diffusion coefficient $D_{2}(j)/D_{0}$ as function of angular momentum
$j$, where $D_{0}=2\protect\Tcoh\nu_{j}^{2}\left(j=0\right)$, for
the different noise models and a specific value of $j_{0}$ (vertical
line). As $j$ decreases, the diffusion coefficients of the correlated
noises deviate from that of the Markovian limit (uncorrelated noise),
and decrease with $j$. The decrease is sharper the smoother the noise.}
\end{figure}

This effective FP equation has the same form and steady state solution
as the one obtained in the Markovian limit, but with a DC that depends
on the noise model through its correlation function. Generally, the
properties of $D_{2}$ will depend on the ``smoothness'' of the
noise, as reflected by the behavior of the ACF at $t\to0$ (Figure
\ref{fig:DCs}). For example, when the noise is generated by an Ornstein\textendash Uhlenbeck
process \cite{Uhlenbeck1930}, the noise is continuous but not differentiable,
and therefore the ACF is not differentiable at $t=0$, and has the
form $C\left(t\right)=e^{-|t|/\Tcoh}$. The DC is then 
\begin{equation}
D_{2}\left(j\right)=2\Tcoh\nu_{j}^{2}\left(j\right)/\left(1+\Tcoh^{2}\nu_{p}^{2}\left(j\right)\right).
\end{equation}

However, such a non-differentiable ACF cannot provide an exact description
of the dynamics. The physical noise is generated by the orbital motion
of the background stars, and is therefore inherently smooth, with
an infinitely differentiable ACF at $t=0$. The smoothness implies
that all the derivatives, $\mathrm{d^{n}}C\left(s\right)/\mathrm{d}t^{n}|_{t=0}=\left(2\pi\right)^{-1}\int_{-\infty}^{+\infty}(i\nu)^{n}S_{\eta}(\nu)\mathrm{d}\nu$,
are finite (all odd orders are zero by symmetry), so that $S_{\eta}(\nu)$
must fall faster than polynomial. That is, the integrated power of
the noise is negligible beyond some characteristic frequency $\nu_{\max}$.
When $\nu_{p}>\nu_{\max}$, there is little power in the noise at
the precession frequency, and the evolution of $j$ decouples from
the background, as is reflected by the diffusion coefficient, $D_{2}\propto S_{\eta}(\nu_{p})=\mathcal{F}\left[C\left(t\right)\right](\nu_{p})$
(Eq. \ref{eq:D_a}). Conversely, the Fourier transform of a non-differentiable
noise (with a non-differentiable ACF at $t=0$) contains all frequencies,
and in particular has a non-vanishing power at all precession frequencies,
which implies a net, noise-driven evolution of $j$ across all phase-space.
The sharp suppression of diffusion when $\nu_{p}>\nu_{\max}$ $ $
is a manifestation of the general concept of adiabatic invariance,
where the action conjugate to a fast dynamical angle is approximately
conserved under slow parametric changes of the Hamiltonian \cite{lan+76}. 

Generally, a physically valid ACF must have the perturbative form
of $C\left(t\right)=1-\frac{1}{2}\left\langle \dot{\eta}_{i}^{2}\left(t\right)\right\rangle t^{2}+{\cal O}(t^{4})$,
where $\left\langle \dot{\eta}_{i}^{2}\left(t\right)\right\rangle \propto\nu_{\max}^{2}$
is the variability frequency of the noise\footnote{For example, for a white noise model truncated at $\nu_{\max}$, $T_{\mathrm{coh}}=\pi/2\nu_{\max}$,
$\left\langle \dot{\eta}_{i}^{2}\left(t\right)\right\rangle =\ddot{-C(0)}=\nu_{\max}^{2}/3$.}, which is not necessarily related to $\Tcoh$. The angular momentum
scale $j_{0}$, where adiabatic invariance becomes important, is therefore
given by $\nu_{p}(j_{0})\sim\nu_{\max}$. In the context of relativistic
dynamics, and noise models with a single scale $\Tcoh\propto1/\nu_{\max}$,
which are considered here for simplicity, it is given by $\nu_{GR}/j_{0}^{2}=2\pi/\Tcoh$,
or 
\begin{equation}
j_{0}=\sqrt{\Tcoh\nu_{GR}/2\pi}\,.\label{eq:j0}
\end{equation}
For example, the Gaussian ACF, $C\left(t\right)=\exp\left(-\pi\left(t/\Tcoh\right)^{2}/4\right)$,
where $\Tcoh=\sqrt{\pi/2\left\langle \dot{\eta}_{i}^{2}\left(t\right)\right\rangle }$,
corresponds to a smooth noise with a DC 
\begin{equation}
D_{2}\left(j\right)=2\Tcoh\nu_{j}^{2}\left(j\right)\exp\left(-\Tcoh^{2}\nu_{p}^{2}\left(j\right)/\pi\right),\label{eq:D2_gauss}
\end{equation}
which decays very strongly as the precession period falls below the
coherence time. 

Different noise models and their corresponding ACFs and DCs of the
$l=1$ relativistic Hamiltonian $H_{\eta}^{GR}$ (Eq. \ref{eq:HetaGR}),
where $\nu_{p}(j)=\nu_{GR}/j^{2}$ and $\nu_{j}\propto\sqrt{1-j}$,
are shown in Figure \ref{fig:DCs}. The results illustrate the relations
between the smoothness of the noise, the continuity of the ACF at
$t=0$ and the suppression of the DC. The implications of this suppression
to the dynamics of relativistic orbits described by $H_{\eta}^{GR}$
are discussed in detail in Section \ref{sec:results}. 

\makeatletter{}
\section{A relativistic induced barrier in phase space}

\label{sec:results}

The dynamics of a test star in a stellar cusp around a MBH depend
on the dynamics of the background stars, described here in terms of
the ACF of a stochastic noise. We analyze here the implications of
two specific noise models on the dynamics of the $l=1$ relativistic
Hamiltonian (Eq. \ref{eq:HetaGR}). These represent two limits, that
of a continuously differentiable noise, with a Gaussian ACF and that
of a non-differentiable noise, with an exponential ACF.

\subsection{Time evolution of the cumulative distribution function}

\label{sub:FPEOM}

\begin{figure}
\includegraphics[width=0.35\paperwidth]{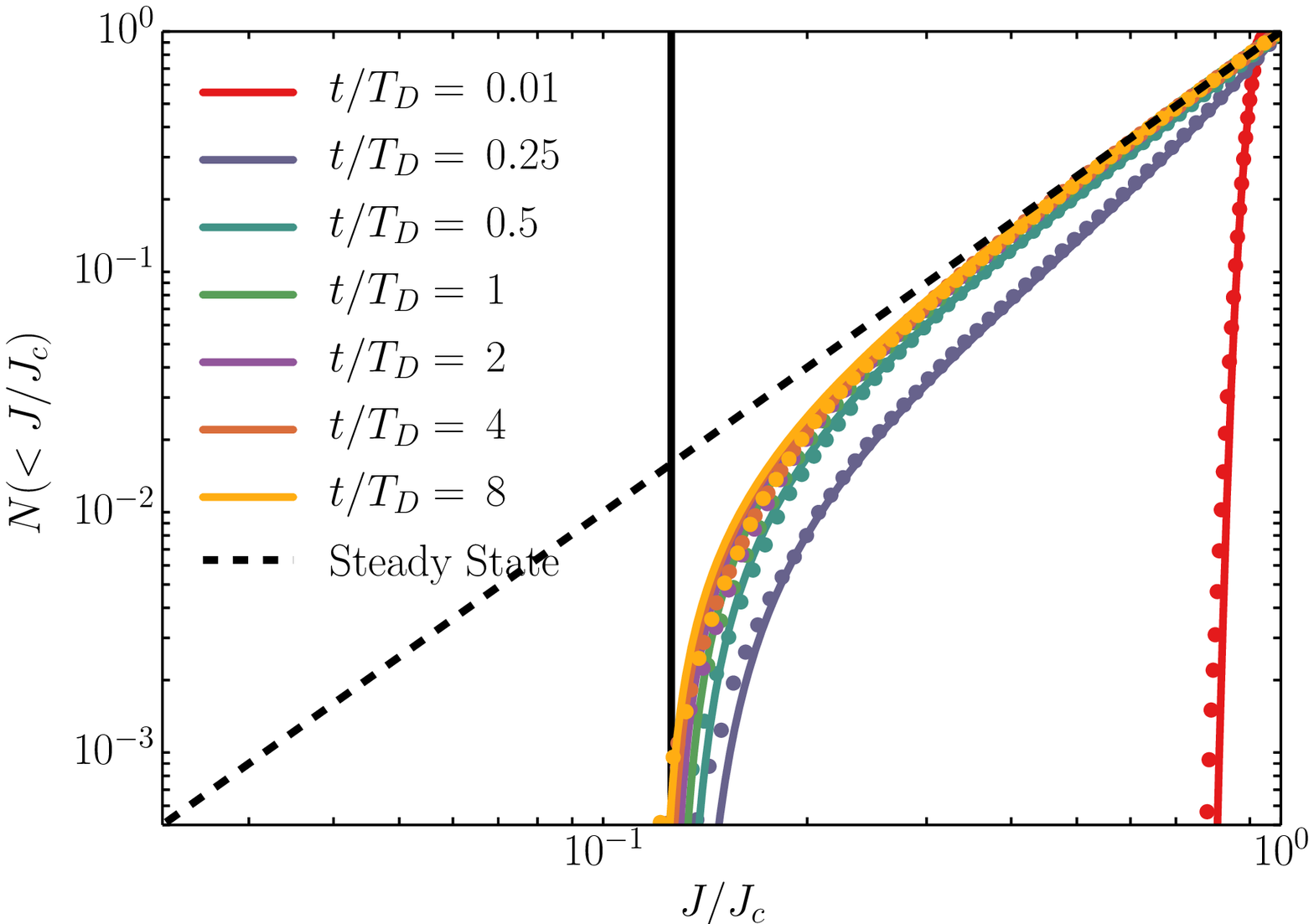}\includegraphics[width=0.35\paperwidth]{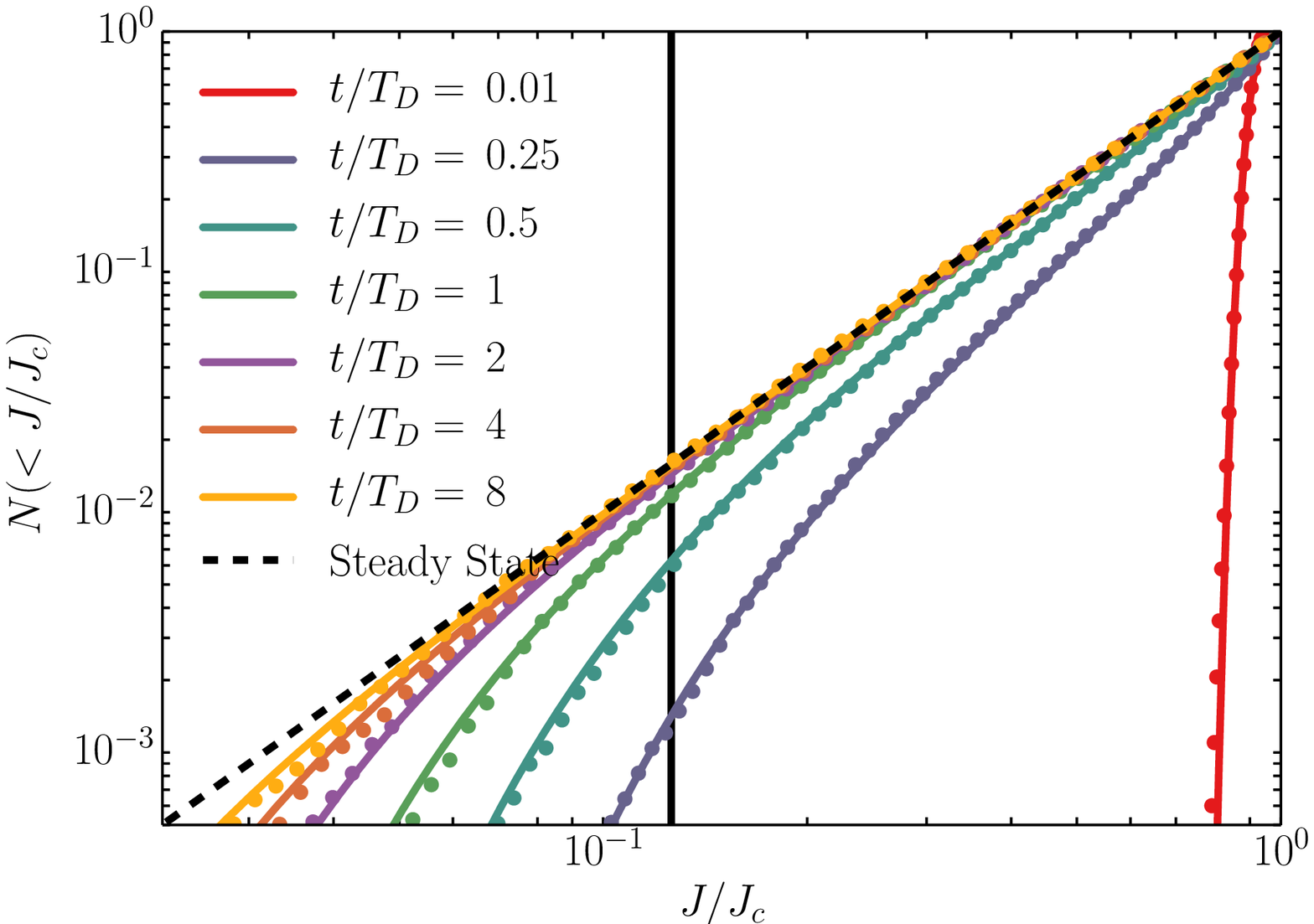}\protect\caption{\label{fig:cdf}Evolution of the cumulative distribution function
with time for the Gaussian ACF noise (left) and for the exponential
ACF noise (right) for a specific value of $j_{0}$ and $T_{c}=0.01T_{D}$.
The integration of the FP equation (Eq. \ref{eq:EffFP}) (solid lines)
agrees with results from the stochastic EOM (Eqs. \ref{eq:EOMJ},
\ref{eq:EOMpsi}) with randomly generated noise sequences (circles).
For Gaussian ACF noise, the system reaches a quasi-steady state which
drops rapidly at $j_{0}$ (vertical line). For exponential ACF noise,
the system approaches steady state (dashed line) in one diffusion
timescales $T_{D}=\nu_{j}^{-2}\left(0\right)/\protect\Tcoh$. }
\end{figure}

\begin{figure}
\centering{}\includegraphics[width=0.4\paperwidth]{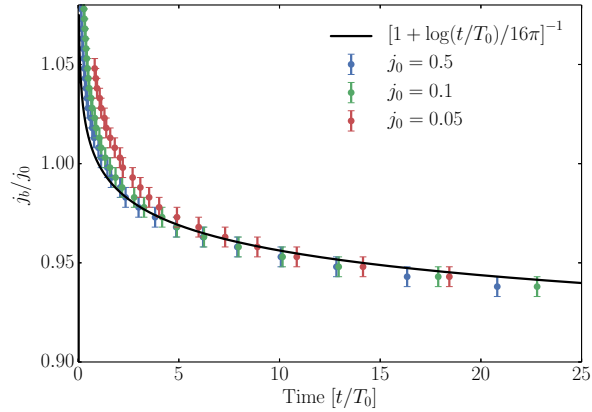}\protect\caption{\label{fig:front}Evolution of the barrier location, $j_{b}(t)$,
of the quasi-steady state probability distribution function under
a Gaussian ACF noise for different values of $j_{0}$, where $j_{0}=j_{b}\left(T_{0}\right)$
and $T_{0}\approx120j_{0}^{2}T_{D}$. The FP results (circles) are
compared to an analytically approximated logarithmically-suppressed
evolution function (solid line) (see \ref{sec:Appendix}). }
\end{figure}

Figure \ref{fig:cdf} shows the evolution of the cumulative distribution
function (CDF) at logarithmic time intervals for the two noise models.
Time is measured in terms of the global diffusion timescale, $T_{D}=\nu_{j}^{-2}\left(j=0\right)/\Tcoh$,
that is, the inverse of the DC in the Markovian limit. The CDF was
obtained by two methods:
\begin{enumerate}
\item Numerical integration of the effective FP equation (\ref{eq:EffFP}),
starting from a narrow PDF (reflecting the initial scatter on timescale
$\Tcoh$, see below) centered around $j_{i}=0.9$ and zero-flux boundary
conditions at $j=1$ and at $j_{\min}\to0$.
\item Numerical integration of the EOM\footnote{Using the \textsc{vode}\_\textsc{f90} \cite{vodef90} implementation
of the \textsc{vode} solver \cite{Brown1989}.} (Eqs. \ref{eq:EOMJ}, \ref{eq:EOMpsi}) are carried out for many
realizations of the noise in a Monte-Carlo fashion. Continuous noise
realizations were generated from the ACF using a discrete Fourier
transform with randomly drawn phases (e.g. \cite{gar+99}). The simulations
were started with an initial $j_{i}=0.9$ and a random orientation
of the orbit. Several values of $\Tcoh\in\left[0.01,1\right]$ and
$\nu_{GR}\in\left[0.1,1\right]$ were used to confirm that the evolution
of the CDF depends only on $j_{0}=\sqrt{\Tcoh\nu_{GR}/2\pi}$. 
\end{enumerate}
Figure \ref{fig:cdf} shows that the dynamics of the system crucially
depend on the noise model. Under the Gaussian ACF (differentiable)
noise, the rate of evolution of the PDF at $j<j_{0}$ is exponentially
suppressed since the local diffusion time diverges as $D_{2}^{-1}(j)\propto\exp\left[-4\pi\left(j_{0}/j\right)^{4}\right]$.
When the stars are initially placed above $j_{0}$, the PDF reaches
a quasi-steady state on a timescale of $T_{D}$, where it drops rapidly
to zero at $j_{b}(t)\lesssim j_{0}$. Figure \ref{fig:front} shows
that the effective boundary of the distribution $j_{b}\left(t\right)$,
defined as the maximum of $dP\left(j,t\right)/dj$, is effectively
constant in time, since as we show in \ref{sec:Appendix}, $j_{b}(t)\approx\left[1+\log\left(t/T_{0}\right)/16\pi\right]^{-1}$
where $j_{b}\left(T_{0}\right)=j_{0}$ and $T_{0}\approx120j_{0}^{2}T_{D}$.
The locus $j_{0}(a)$ can then be interpreted as an effective barrier
in phase space, that is, stellar trajectories that cross $j_{0}$
from above, spend only a vanishingly small time below it. In marked
contrast, under the exponential ACF (non-differentiable) noise, the
PDF evolves much faster and is suppressed only as $D_{2}^{-1}(j)\propto\left(j_{0}/j\right)^{4}$.
It does not display a definite barrier, and the fall of the PDF at
low-$j$ merely reflects the finite time of the simulation (Figure
\ref{fig:cdf}). Note for comparison that the rate of evolution under
uncorrelated (discontinuous) noise, equivalent to the case of no GR
precession, is nearly uniform in $j$ for $j<1$ (Figure \ref{fig:DCs}),
and therefore evolves even faster.


\subsection{Evolution of low-$j$ orbits}

\label{sub:lowj}

\begin{figure}
\includegraphics[width=0.35\paperwidth]{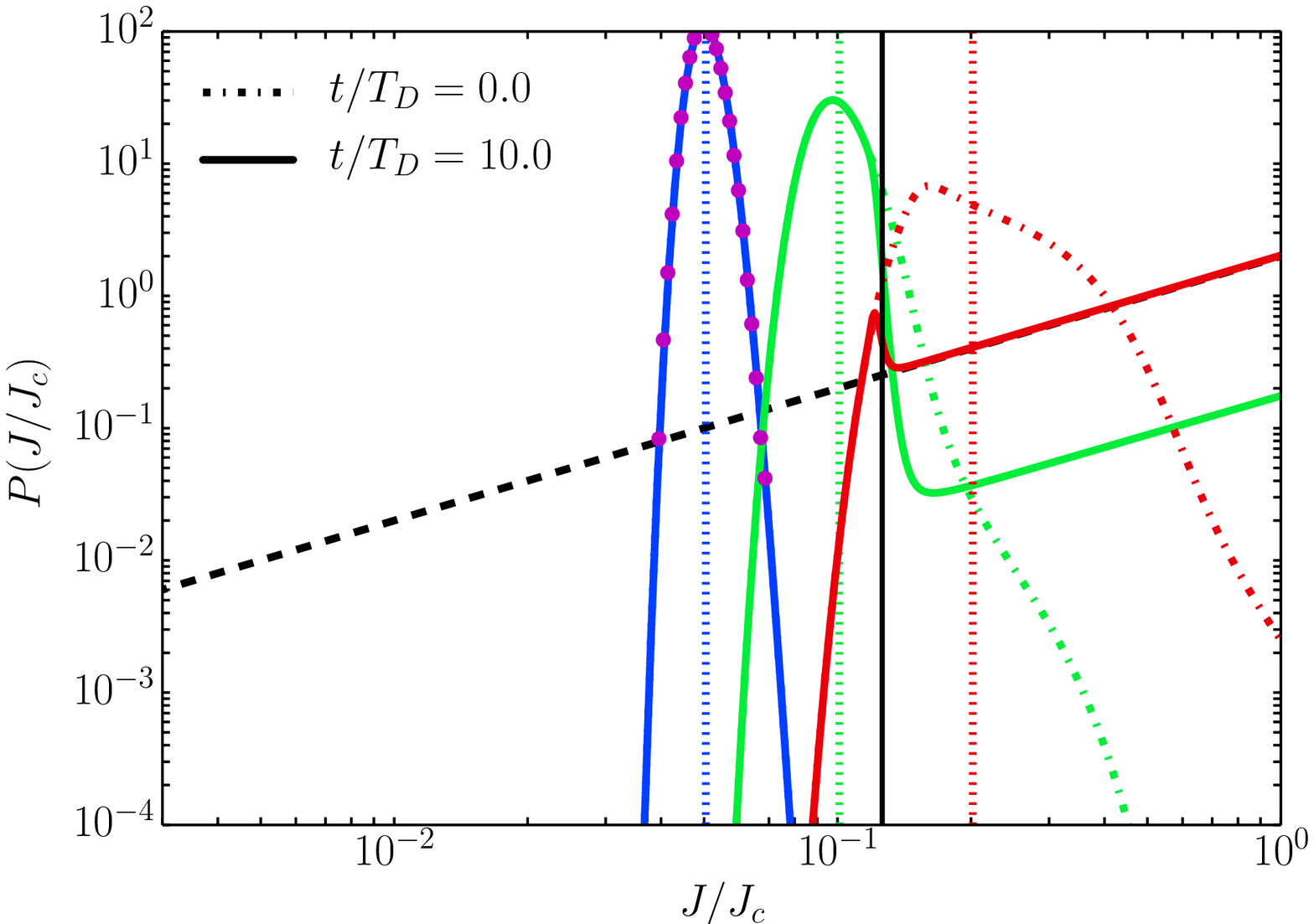}\includegraphics[width=0.35\paperwidth]{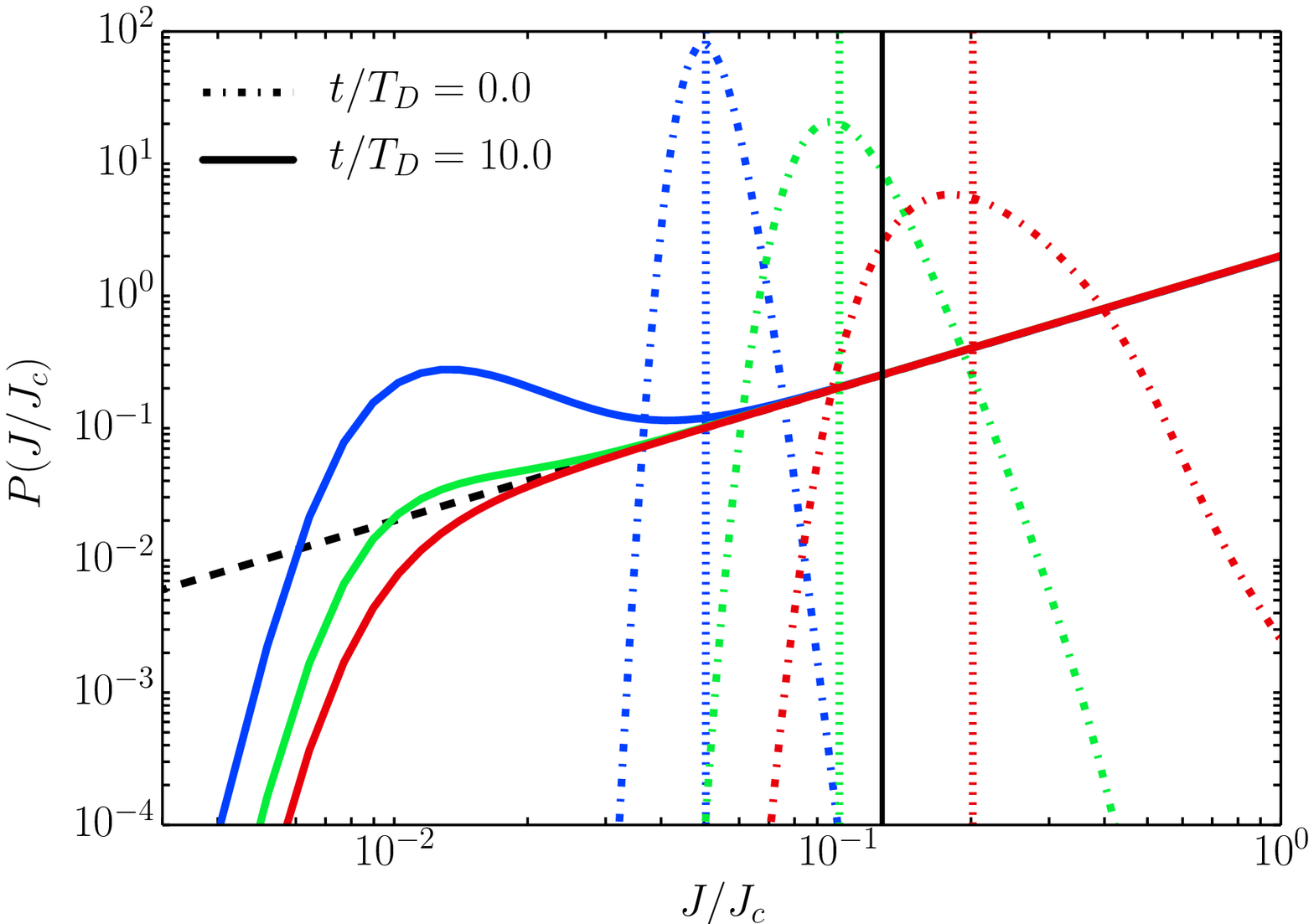}\protect\caption{\label{fig:sub_bar}Evolution of low-$j$ orbits for the Gaussian
ACF noise (left) and for the exponential ACF noise (right), for a
given $j_{0}$. Starting with an initial PDF (dash-dotted lines) centered
around $j_{i}$ (vertical dotted lines) we evolve the system using
the FP equation to $t=10T_{D}$ (solid lines). At $t=0$ the PDF reflects
the coherent evolution phase. Stars starting (after the coherent evolution
phase) with initial $j$ above $j_{0}$ (vertical solid line) converge
to their steady state (dashed-line) in a finite time. The corresponding
results from the integration of the stochastic EOM are shown for the
Gaussian ACF noise, for the model with lowest $j_{i}$ at $t=10T_{D}$
(circles). }
\end{figure}

We integrated the system starting with an initial $j_{i}$ close to
$j_{0}$. Figure \ref{fig:sub_bar} shows that the evolution of the
PDF strongly depends on the noise model and on the location of $j_{i}$.
For the Gaussian ACF (differentiable) noise, test stars with initial
$j_{i}\gg j_{0}$ reach the equilibrium maximal entropy distribution
by time $t\approx T_{D}$, while stars starting at $j_{i}\ll j_{0}$
remain near $j_{i}$, out of equilibrium, on times $t\gg T_{D}$.
We model here the evolution from these two opposite regimes and compare
results from the FP equation and direct integration of the stochastic
EOM. This comparison must however take into account the fact that
the FP is applicable only for $t>\Tcoh$, while the directly integrated
orbits evolve on short timescales. We match the initial conditions
of the two methods by choosing the initial PDF of the FP equation
to have a specific shape and width that reflects the evolution of
the integrated orbits up to $t\sim\Tcoh$. An ensemble of stars with
the same initial $j_{i}$ will rapidly spread in $j$ after a short
time $t<\Tcoh$, while the noise is almost constant and $j$ evolves
linearly. The evolution of each realization will depend on the specific
initial values of the test star's orbital orientation and the value
of the noise . To estimate the PDF at $t=\Tcoh$, we define $g=j_{i}/j-1$
and use the fact that on timescale $t\ll\Tcoh$, the Hamiltonian $H_{\eta}^{GR}$
(\ref{eq:HetaGR}) is time independent. Thus 
\begin{equation}
g\approx j_{i}\frac{\nu_{j}\left(j_{i}\right)}{\nu_{GR}}\left[\hat{e}_{\psi}(\boldsymbol{\Pi})-\hat{e}_{\psi}(\boldsymbol{\Pi}_{i})\right]\cdot\boldsymbol{\eta}
\end{equation}
which implies that $g$ is a random Gaussian with zero mean and $\sigma_{g}^{2}\approx2j_{i}^{2}\left(\nu_{j}\left(j_{i}\right)/\nu_{GR}\right)^{2}\left(1-\left\langle \cos\left(\psi_{i}-\psi\right)\right\rangle \right)$
where we assume that $j\approx j_{i}$ and $\Euler\approx\Euler_{i}$$ $.
For $t\gg j^{2}/\nu_{GR}$, the phase $\psi$ is randomized and $\left\langle \cos\left(\psi_{i}-\psi\right)\right\rangle \approx0$.
For $t\ll j^{2}/\nu_{GR}$, $\psi\approx\psi_{i}+\nu_{GR}t/j^{2}$
and $\left\langle \cos\left(\psi_{i}-\psi\right)\right\rangle \approx1-\nu_{GR}^{2}t^{2}/\left(2j^{4}\right)$.
Therefore we can estimate the $j$ distribution after the coherent
phase $t\sim\Tcoh$ by
\begin{equation}
P_{0}\left(j;j_{i},\sigma_{g}\right)=\frac{1}{j^{2}}\frac{1}{\sqrt{2\pi\sigma_{g}^{2}}}\exp\left[-\left(j-j_{i}\right)^{2}/\left(2\sigma_{g}^{2}j^{2}\right)\right],\label{eq:P0}
\end{equation}
with
\begin{equation}
\sigma_{g}^{2}=\frac{\nu_{j}^{2}\left(j_{i}\right)\Tcoh^{2}}{j_{i}^{2}}\left\{ \begin{array}{cc}
j_{i}^{4}/\left(2\pi^{2}j_{0}^{4}\right) & j_{i}\le\left(2\pi^{2}\right)^{1/4}j_{0}\\
1 & j_{i}>\left(2\pi^{2}\right)^{1/4}j_{0}
\end{array}\right..
\end{equation}
$P_{0}(j;j_{i},\sigma_{g})$ therefore provides the effective initial
conditions that correctly match those of the stochastic EOM, as verified
by the results shown in Figures \ref{fig:cdf} and \ref{fig:sub_bar}.

\makeatletter{}

\section{Discussion\label{sec:Discussion}}

The description of the dynamics near a MBH involves many processes,
such as two-body relaxation, resonant relaxation, GR-precession and
mass precession. Direct relativistic $N$-body simulations include,
in principle, all these effects, but are very hard to interpret since
these complex dynamics are entangled, and the computational costs
limit the simulations to small $N$.

In this study we addressed key elements of this problem analytically,
and focused on the interplay between the deterministic GR precession
and the stochastic fluctuations of the background torques. We neglected
stochastic two-body energy and angular momentum relaxation, since
these operate on much longer timescales. We also omitted precession
due to the extended stellar mass around the MBH, since it is negligible
in the $j\to0$ limit that is particularly relevant for relativistic
orbits.

We demonstrated how the complicated dynamics of a nearly-Keplerian
$N$-body system can be described and studied in a formal statistical
mechanics framework. We described the dynamics of a test star by a
stochastic orbit-averaged Hamiltonian, where the noise terms represent
the time-dependent evolution of the background. We extended the usual
Legendre multipole expansion in real space to include the new degree
of freedom that is introduced by orbit-averaging, namely the orientation
of the Keplerian ellipse in its plane, by expanding the Hamiltonian
in Wigner matrices. We then derived the explicit stochastic 3D EOM
from the first order ($l=1$) relativistic Hamiltonian $H_{\eta}^{GR}$,
and presented the corresponding effective FP equation for a general
correlated Gaussian noise. We validated that the numerical integration
of this FP equation reproduces the statistical properties of the stochastic
EOM, and importantly, converges to the maximal entropy limit for $t\to\infty$
(Section \ref{sec:results}).

We showed that evolution toward low-$j$ orbits under the $H_{\eta}^{GR}$
Hamiltonian with Gaussian ACF noise (i.e. smooth noise) is restricted
by adiabatic invariance. Stars that are initially on high-$j$ orbits
where the GR precession time is long, evolve rapidly to lower $j$
and to faster precession, until they reach a threshold at $j_{0}$,
where the precession time falls below the coherence time of the background.
Beyond that point, the probability of finding a star is vanishingly
small, because the fast precession effectively decouples the orbit
from the effects of the noise through the mechanism of adiabatic invariance.
This is formally expressed by the strong suppression of the diffusion
coefficients below $j_{0}$ (steeper than exponential in $j$) (Section
\ref{sub:FPEOM}). The phase-space locus $j=j_{0}(a)$ is not a reflecting
boundary, but a barrier in the sense the steps toward $j_{0}$ become
infinitely small, while steps away from $j_{0}$ become larger. Due
to this asymmetry, stars will spend on average only a short time near
the barrier. However, we demonstrate that if stars start initially
at $j<j_{0}$, for example by being tidally captured there in a binary
disruption event, the diffusion timescale back to $j>j_{0}$ is so
long that they stall near their capture orbit (Section \ref{sub:lowj}).
Note that this result does not take into account the fact that stochastic
2-body relaxation, not included in our treatment here, will eventually
push the star to higher $j$ and therefore to faster evolution. 

This should be contrasted with the behavior of the system under exponential
ACF noise (i.e. non-differentiable noise), where there is no barrier,
and the deviation from steady state reflects only the finite age of
the system (Section \ref{sec:results}; Figure \ref{fig:cdf}). Since
the system evolves toward isothermal steady state ($P(j)=2j$), where
most of the stars are at high-$j$, the trajectories of stars that
start out at low-$j$ tend to end up at higher-$j$, while stars that
start out at high-$j$, tend to remain there. Such a behavior was
observed in the Monte Carlo simulation \cite{Antonini2013,Antonini2014}
that used a continuous, but not continuously-differentiable noise
model (see \cite{Merritt2011} for details), which most closely corresponds
to our exponential ACF noise model.

The main limitations of this study are that we considered only the
$l=1$ term in the Hamiltonian and introduced the physical properties
of the noise (smoothness and timescales, i.e. the form of the ACF)
as free parameters. However, the maximal variability frequency of
the noise, and the phase-space locus of the barrier, can be estimated
by general considerations. The noise is a function of the orbital
elements of the background stars (Eq. \ref{eq:etalnm}), which evolve
at a rate that is a combination of the deterministic precession due
to GR, and due to the enclosed mass, and a stochastic precession due
to the residual torques themselves. A physical correlation function
(i.e. differentiable at the origin) can be expanded around $t^{\prime}=0$
by 
\begin{equation}
\left\langle \eta\left(t\right)\eta\left(t+t^{\prime}\right)\right\rangle =1-\frac{1}{2}\left\langle \dot{\eta}^{2}\right\rangle t^{\prime}{}^{2}+{\cal O}\left(t^{\prime}{}^{4}\right)\,.
\end{equation}
Recall that the barrier phenomenon is related to the maximal variability
frequency of the noise (Section \ref{sub:nonMarkov}). Assume that
the fastest precession rate of a typical (i.e. not particularly eccentric)
background star is due to mass precession, $\nu_{M}(a)\propto N(a)$
(ignoring GR precession of the background stars). This deterministic
in-plane precession is typically faster by $\nu_{M}/\nu_{j}\propto\sqrt{N}$
(Eqs. \ref{eq:HetaGR}, \ref{eq:nuj}--\ref{eq:nupsi}) than the stochastic
evolution of the other orbital elements \cite{Hopman2006}, and it
therefore dominates the noise evolution rate $\sqrt{\left\langle \dot{\eta}^{2}\right\rangle }$.
To estimate its magnitude, we note that the torque on a test star
with semi-major axis $a$ is dominated by contributions from background
stars at $\sim2a$ \cite{Gurkan2007}, and the relevant background
precession rate is $\sqrt{\left\langle \dot{\eta}^{2}\right\rangle }\approx\nu_{M}\left(2a,e_{m}\right)$
where here the median eccentricity $e_{m}=\sqrt{1/2}$ was taken as
a characteristic value. Under these assumptions, for a Gaussian ACF
$\Tcoh=\sqrt{\pi/2\left\langle \dot{\eta}^{2}\right\rangle }\approx\sqrt{\pi/2}\nu_{M}^{-1}$,
and the barrier is located at $j_{b}^{2}=\nu_{GR}\nu_{M}^{-1}/\sqrt{8\pi}\propto1/\left(N\left(a\right)a\right)$.
Note that this scaling is different from a previous empirical determination
of $j_{b}\approx0.6\left(a/1\mathrm{mpc}\right)^{-3/2}$, which was
based on a qualitative fit to $N$-body results for an isothermal
($N(a)\propto a$) stellar cusp model \cite{Merritt2011,Brem2013}.
For that system, our analysis\footnote{For an $\alpha=2$ cusp with noise described by a Gaussian ACF and
assuming $\Tcoh=\sqrt{\pi/2}\nu_{M}^{-1}\left(2a,\sqrt{1/2}\right)$
, $j_{b}^{2}\simeq0.4(r_{g}/a)(Q/N(a))$. Alternatively, assuming
$\Tcoh=Q\nu_{r}^{-1}\left(a\right)/\sqrt{N(2a)}$, $j_{b}^{2}\simeq0.33\left(r_{g}/a\right)\left(Q/\sqrt{N\left(a\right)}\right)$} indicates that the barrier should be at  $j_{b}\approx0.6\left(a/1\mathrm{mpc}\right)^{-1}$.
Alternatively, if the coherence time of the noise reflects the stochastic
evolution of the background stars due to the residual torque themselves,
then $\Tcoh\approx Q\nu_{r}^{-1}\left(a\right)/\sqrt{N(2a)}$, and
an even flatter relation $j_{b}\approx0.4a^{-3/4}$ is predicted.
Either loci appear to be reasonably consistent with the presently
available numeric data, which has a limited range in $a$ and $j$.
Recently, Hamers et al. \cite{Hamers2014}, measured the diffusion
coefficient from a suite of simulations. Based on these diffusion
coefficients, they presented two expression for the barrier location.
One is consistent with \cite{Merritt2011} and the other is consistent
with the analysis presented here. However, using the available data
they could not rule out either one. Therefore, formal fits and better
statistics will be required to discriminate between the different
models. 

It should be emphasized that the neglected higher ($l>1$) terms in
the multipole expansion of the Hamiltonian, while smaller in magnitude,
could in principle introduce shorter timescales in the noise, and
correspondingly result in diffusion coefficients that decay at smaller
$j$. The superposition of many such decaying terms could result in
a slower, power-law decay, which will blur the barrier. Further study
is required to apply the insights obtained here about dynamics driven
by correlated noise, and in particular the close relation between
the smoothness of the background noise, adiabatic invariance, and
very low-angular momentum orbits, to real physical systems.

\ack{}{}

We acknowledge support by the ERC Starting Grant No. 202996, DIP-BMBF
Grant No. 71-0460-0101, under which this project was initiated, and
the I-CORE Program of the PBC and ISF (Center No. 1829/12).

\appendix

\makeatletter{}
\section{Time evolution of the barrier for a Gaussian ACF noise}

\label{sec:Appendix}

The location of the barrier $j_{b}\left(t\right)$ can be defined
as the maximum of $\partial P\left(j,t\right)/\partial j$. Assuming
that for $j\approx j_{b}$, $P\left(j,t\right)$ is a self-similar
function of $j-j_{b}$, that is $P\left(j,t\right)\approx P\left(j-j_{b}\left(t\right)\right)$,
and $\int_{j_{b}}^{1}P\left(j,t\right)dj$ is constant in time, we
obtain
\begin{eqnarray}
P\left(j_{b}\left(t\right),t\right)\dot{j}_{b}\left(t\right) & = & \int_{j_{b}\left(t\right)}^{1}\dot{P}\left(j,t\right)dj,
\end{eqnarray}
and
\begin{equation}
\dot{j}_{b}\left(t\right)\frac{\partial}{\partial j}P\left(j_{b}\left(t\right),t\right)=-\dot{P}\left(j_{b}\left(t\right),t\right).
\end{equation}
Using Eq. (\ref{eq:EffFP}) and $\partial^{2}P\left(j_{b}\left(t\right),t\right)/\partial j^{2}=0$
we obtain
\begin{eqnarray}
\dot{j}_{b}\left(t\right) & = & D_{2}(j_{b})\frac{1}{j_{b}}-\frac{1}{2}D_{2}^{\prime}(j_{b})\label{eq:jbdot}
\end{eqnarray}

For the Gaussian ACF assuming $\nu_{p}=\nu_{GR}/j^{2}$ the DC is
given by (Eq. \ref{eq:D2_gauss}), 
\begin{equation}
D_{2}\left(j\right)=2T_{D}^{-1}\left(1-j\right)e^{-4\pi\left(j_{0}/j\right)^{4}},
\end{equation}
where $T_{D}=\nu_{j}^{-2}\left(j=0\right)/\Tcoh$. By defining $x=j_{0}/j_{b}$
and $s=t/T_{D}$, we use Eq. (\ref{eq:jbdot}) to obtain 
\begin{eqnarray}
\frac{dx}{ds} & = & j_{0}^{-2}\left[\left(16\pi x^{4}-1\right)\left(x-j_{0}\right)-x\right]x^{2}e^{-4\pi x^{4}}\nonumber \\
 & \approx & j_{0}^{-2}\left[\left(16\pi-1\right)\left(1-j_{0}\right)-1\right]e^{4\pi\left(3-4x\right)}.
\end{eqnarray}
where we assumed $x\approx1$. Therefore, as demonsrated in Figure
\ref{fig:front}, the evolution of the barrier can be approximated
by a logarithmically-suppressed evolution function

\begin{equation}
\frac{j_{b}\left(t\right)}{j_{0}}=\left[1+\frac{1}{16\pi}\log\left(t/T_{0}+e^{16\pi\left(j_{0}/j_{b}\left(0\right)-1\right)}\right)\right]^{-1}\approx\left[1+\frac{1}{16\pi}\log\left(t/T_{0}\right)\right]^{-1},
\end{equation}
where 
\begin{equation}
T_{0}=\frac{e^{4\pi}}{16\pi}\frac{j_{0}^{2}}{\left(16\pi-1\right)\left(1-j_{0}\right)-1}T_{D}\approx120j_{0}^{2}T_{D},
\end{equation}
is the time the barrier reaches the point $j_{b}\left(t\right)=j_{0}$.
Note that since $j_{0}^{2}T_{D}=\nu_{GR}\nu_{j}^{-2}\left(j=0\right)/2\pi$,
$T_{0}$ is independent of $\Tcoh$.

\bibliographystyle{unsrt}
\bibliography{ms}

\end{document}